\documentstyle[aps,multicol,epsfig]{revtex}
\begin{document}
\title{Motion of vortex lines in quantum
mechanics\footnote{To appear in Physical Review A.}}
\author{Iwo Bialynicki-Birula\footnote{email: birula@cft.edu.pl}$^{1,4}$,
Zofia Bialynicka-Birula$^{2,3}$, and Cezary \'Sliwa$^{1,3}$}
\address{Center for Theoretical Physics$^1$,
Institute of Physics$^2$, College of Science$^3$,\\ Al. Lotnik\'ow
32/46, 02-668 Warsaw, Poland and\\ Institute of Theoretical
Physics$^4$, Warsaw University, Ho\.za 69, 00-681 Warsaw, Poland}
\maketitle
\begin{abstract}
Exact analytic solutions of the time dependent Schr\"odinger equation
are produced that exhibit a variety of vortex structures. The
qualitative analysis of the motion of vortex lines is presented and
various types of vortex behavior are identified. Vortex creation and
annihilation and vortex interactions are illustrated in the special
cases of the free motion, the motion in the harmonic potential, and in
the constant magnetic field. Similar analysis of the vortex motions is
carried out also for a relativistic wave equation.
\end{abstract}
\begin{multicols}{2}

\section{Introduction}

Vortices have been a source of fascination since time immemorial.
Empedocles, Aristotle \cite{arist}, and Descartes \cite{desc} tried to
explain the formation of the Earth, its gravity, and the dynamics of
the whole solar system as due to primordial cosmic vortices. Lord
Kelvin \cite{lord} envisaged atoms as vortex rings. Modern science of
hydrodynamic vortices began with the fundamental work of Helmholtz
\cite{helm} on the theory of vorticity. The present knowledge of vortex
dynamics in fluid mechanics is best summarized in a recent monograph by
Saffman \cite{saff}.

In our paper we shall be concerned with vortices not in ordinary
fluids but in a peculiar fluid --- the probability fluid --- that
can be associated with the wave function of a quantum system.
Vortices of that type have been studied in the past in connection
with superfluidity, superconductivity, and recently in connection
with Bose-Einstein condensation \cite{tt,ds,rok,duan}. The main
tool in the study of these phenomena is the Ginzburg-Landau
(Gross-Pitaevskii) equation. The nonlinear character of this
equation greatly complicates the analysis of time evolution of the
vortices, especially in three dimensions. There are no exact
analytic solutions of this equation with nontrivial vortex
structure. One must resort either to approximations which can
reveal only some features of the vortex motion (as, for example,
in the early work of Fetter \cite{fetter1,fetter2}) or to
extensive numerical calculations (cf., for example, \cite{abk}).
In the present paper we use the standard time-dependent
Schr\"odinger equation to investigate the motion of vortex lines
embedded in the probability fluid of a quantum particle. The
linearity of this equation enables us to obtain a variety of
exact, time-dependent analytic solutions for the wave functions
that describe the probability flow with one, two or more moving
vortex lines. We are also able to study the motion of vortex lines
in the presence of external forces. Our analysis of many special
cases makes it possible to discover certain general phenomena such
as the vortex switchover and the vortex creation and annihilation.
The important conclusion of our analysis is that quantum effects
play a crucial role in those phenomena. We believe that our
analysis captures the essential features of quantum vortex
dynamics that are common to many different situations. In
particular, our analysis also extends to the relativistic domain.

We have to admit that vortex lines associated with wave functions are
rather elusive objects. They might be connected with physical reality
in two ways. First, the vortex lines can appear only at those places
where the wave function is equal to zero; the vanishing of the
probability density definitely has observable consequences. Second, the
wave function of a macroscopically populated state of many bosons
acquires a direct physical interpretation; vortex lines in this case
may become directly observable as, for example, in BEC \cite{mzw,gwm}.

\section{Vortex lines in the hydrodynamic formulation of wave
mechanics}\label{hydro_sec}

In quantum theory vortex lines arise in the hydrodynamic formulation of
wave mechanics. This formulation originally due to Madelung \cite{mad}
(see also a review article \cite{gd}) employs the hydrodynamic
variables to describe the flow of quantum mechanical probability for a
single particle. The probability density $\rho({\bf r},t)$ and the
velocity field ${\bf v}({\bf r},t)$ are connected with the wave
function $\psi({\bf r},t)$ through the formulas
\begin{eqnarray}\label{psi}
  \psi({\bf r},t) = R({\bf r},t)\exp[i S({\bf r},t)/\hbar],
\end{eqnarray}
\begin{eqnarray}\label{hydro1}
  \rho({\bf r},t ) = \vert\psi({\bf r},t)\vert^2 = R^2({\bf r},t),
  \end{eqnarray}
\begin{eqnarray}\label{hydro2}
  {\bf v}({\bf r},t) &=& \frac{1}{m}\frac{{\rm Re}
  [\psi^*({\bf r},t)(-i\hbar\nabla - e{\bf A})\psi({\bf r},t)]}
  {\vert\psi({\bf r},t)\vert^2}\nonumber\\
   &=& \frac{1}{m}(\nabla S({\bf r},t) - e{\bf A}({\bf r},t)),
\end{eqnarray}
where m is the mass of the particle and ${\bf A}({\bf r},t)$ is
the electromagnetic vector potential. The gradient term $\nabla S$
in (\ref{hydro2}) does not contribute to vorticity. Thus, the {\em
bulk} vorticity $\nabla\times{\bf v}({\bf r},t)$ in the
probability fluid is completely determined by the magnetic field
except at the points where the phase $S$ is singular. This may
occur only at the points where the wave function vanishes. The
vanishing of the complex wave function, in turn, gives two
conditions: the vanishing of two real functions. Each condition
defines a two-dimensional surface. The intersection of these two
surfaces defines a line. Thus, in addition to the given
distribution of vorticity in the bulk of the probability fluid,
that is uniquely determined by the magnetic field, we may also
have isolated vortex lines embedded in the fluid. Along these
lines the vorticity has a two-dimensional Dirac delta-function
singularity in the plane perpendicular to each vortex line. It is
the dynamics of these isolated vortex lines that will be the
subject of our study. In this context we shall identify vortex
lines with line vortices.

To ensure that the wave function is single valued, the strength of
every vortex as measured by the circulation $\Gamma$
along any closed contour~$C$ encircling the vortex line,
\begin{eqnarray}\label{circ}
  \Gamma = \oint_C d{\bf l}\!\cdot\!{\bf v}({\bf r},t),
\end{eqnarray}
must be quantized:
\begin{eqnarray}\label{quant}
  \Gamma = \frac{2\pi\hbar}{m}\;n,\,\,\,\,n = 0,\pm 1,\pm 2,\dots.
\end{eqnarray}
As one approaches the vortex line the velocity ${\bf v}({\bf
r},t)$ must tend to infinity in order to satisfy the quantization
condition (\ref{quant}).

\begin{figure}
  \begin{center}
    \epsfbox{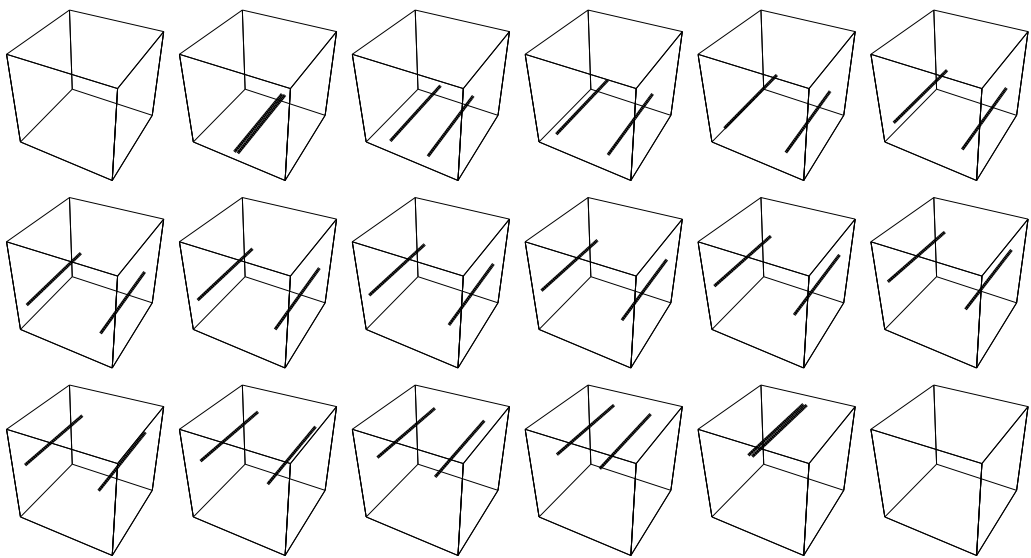}
  \end{center}
\caption{Instantaneous creation and instantaneous annihilation of a
vortex pair with opposite circulation along a line.}\label{fig:vor_creat2}
\end{figure}

The circulation is conserved during the time evolution
\cite{bbbb,bbck} (in full correspondence with the Helmholtz-Kelvin
law for ordinary nonviscous fluids \cite{lamb}). The creation of
vortex lines may proceed through two different scenarios: they may
either appear in the form of a closed vortex line that springs
from a point (Fig.~\ref{fig:vor_creat1}) or they can be created in
pairs of opposite circulation (Fig.~\ref{fig:vor_creat2}).

The hydrodynamic formulation can also be introduced in wave mechanics
of a spinning particle governed by the Pauli equation \cite{bbck,taka,jz}.

Despite all the formal similarities between the hydrodynamics of
classical fluids and the hydrodynamic formulation of quantum
mechanics, there is a fundamental difference between the
properties of line vortices in these two cases. In classical
hydrodynamics, as pointed out by Saffman \cite{saff}, p. 36, ``the
line vortex is a useful kinematic concept, but it does not exist
as a sensible dynamical limit''. In quantum mechanics the line
vortices occur in exact solutions of the Schr\"odinger equation
without any limiting procedure. The fact that the velocity becomes
infinite as one approaches the line vortex is acceptable, because
it is only the probability and not any real matter that flows with
that velocity.

\section{The anatomy of a vortex line}\label{anat}

The commonly known examples of vortex lines are those embedded in the
wave functions that have a fixed value $\hbar m_z$ of the $z$-component
of angular momentum (eg., the spherical harmonics). These vortex lines
run from $-\infty$ to $\infty$ along the $z$-axis and have the
circulation $\hbar m_z/m$. Usually they are stationary, due to angular
momentum conservation. In addition to straight vortex lines, we can
also easily produce stationary vortex rings by superposing wave
functions with the same energy but different angular momenta. The
simplest example is the superposition (with a $\pi/2$ phase difference)
of the electronic $\psi_{200}$ and $\psi_{210}$ states in hydrogen. The
vortex ring is formed in $z=0$ plane and has the radius of two Bohr
radii.

In our study we shall discuss a broader class of vortices that, in
general, do not exhibit cylindrical symmetry even when they form a
straight line. Let us consider an arbitrary vortex line
$\mbox{\boldmath$\xi$}(s)$ parametrized by the length $s$ along the
line. A typical quantum vortex is associated with the line of first
order zeros of the wave function. Higher order zeros will, in general,
lead to vortices that carry more that one unit of vorticity. The local
properties of such a vortex at a point $\mbox{\boldmath$\xi$}(s)$ are
determined by the behavior of the wave function near this point
\begin{eqnarray}\label{taylor}
  \psi(x,y,z) \approx ({\bf r} -\mbox{\boldmath$\xi$}(s))
  \!\cdot\!\nabla\psi(\mbox{\boldmath$\xi$}(s)),
\end{eqnarray}
where we have kept only the lowest terms in the Taylor expansion. Since
quantum vortices have zero thickness, this approximation is always
valid sufficiently close to the vortex line. At each point on the
vortex line let us define a complex vector ${\bf w}(s)$ --- the
gradient of the wave function
\begin{eqnarray}\label{defw}
 {\bf w}(s) = \nabla\psi({\bf r})\vert_{{\bf r}=\mbox{\boldmath$\xi$}(s)}.
\end{eqnarray}
We shall show that this vector describes the basic properties of
the vortex. Let us note that since the derivative of the wave
function along the vortex line with respect to the length
parameter $s$ is zero, the real and the imaginary parts of the
vector ${\bf w}(s)$ lie in the plane perpendicular to the tangent
vector ${\bf t}(s) = d\mbox{\boldmath$\xi$}(s)/ds$,
\begin{eqnarray}\label{perp}
 \frac{d\psi(\mbox{\boldmath$\xi$}(s))}{ds}
  = {\bf t}(s)\!\cdot\!{\bf w}(s) = 0.
\end{eqnarray}
The general complex vector ${\bf w}$ has 4 real components, but only
2 parameters contain the information about the vortex, since the
multiplication of the wave function by a complex number does not
change, of course, the velocity field. Indeed, the velocity field ${\bf
v}(x,y,z)$ calculated for the approximate wave function (\ref{taylor}),
\begin{eqnarray}\label{vel}
  {\bf v}(x,y,z) &=& \frac{\hbar}{2mi}\left(\frac{\bf w}
  {{\bf w}\!\cdot\!{\bf r}}-\frac{{\bf w}^*}
  {{\bf w}^*\!\cdot\!{\bf r}}\right)\nonumber\\
   &=&\frac{\hbar}{2mi}\frac{{\bf r}\times({\bf w}\times{\bf w}^*)}
   {\vert{\bf w}\!\cdot\!{\bf r}\vert^2},
\end{eqnarray}
is a homogeneous function of ${\bf w}$. Therefore, ${\bf v}$ does not
change when ${\bf w}$ is multiplied by any complex number. For
simplicity, we have chosen the origin of the coordinate frame at the
point $\mbox{\boldmath$\xi$}(s)$. Only in the special case, when the
real and imaginary part of ${\bf w}$ are mutually orthogonal and of
equal length, the velocity lines follow circles in the plane orthogonal
to the vortex line. In all other cases the vortex is ``squeezed''; the
velocity lines follow ellipses. In the degenerate case, when the real
part and the imaginary part of ${\bf w}$ are parallel, the ellipse is
squeezed into a line and the velocity disappears. The vector ${\bf w}$
will, in general, vary as we move along the vortex line. However, one
may check by a direct calculation that the circulation (\ref{circ})
calculated for the velocity field (\ref{vel}) does not depend on ${\bf
w}$ and is equal to $\pm2\pi\hbar/m$ (the sign depends on the
orientation of the contour $C$). Higher values of the circulation are
obtained when the wave function has zeros of the higher order but the
typical case, of course, is the first order zero.

The above analysis shows that the appearance of vortex lines in
wave mechanics does not require that some special conditions be
met. On the contrary, a generic wave function has vortex lines.
The vanishing of the real and the imaginary part of the wave
function defines a line. The Taylor expansion of the wave function
around the point lying on this line defines the complex vector
${\bf w}$ that determines the local structure of the vortex.

The vector ${\bf w}$ not only determines the velocity field but also
plays a crucial role in determining the motion of the vortex line
itself. The velocity~${\bf u}=d\mbox{\boldmath$\xi$}(s,t)/dt$ of a
point~$\mbox{\boldmath$\xi$}(s,t)$ lying on the vortex line can be
obtained from the condition
\begin{eqnarray}
d\psi(\mbox{\boldmath$\xi$}(s,t),t)/dt={\bf u}\cdot{\bf w} +
\partial\psi/\partial t = 0.\label{velocity}
\end{eqnarray}
The vector ${\bf u}$ is defined only up to a vector parallel to the
vortex line and it can be determined by solving Eq.(\ref{velocity}):
\begin{eqnarray}
  {\bf u} = \frac{{\bf w}\times{\bf w}^*}{\vert{\bf w}\times{\bf w}^*\vert^2}
    \times\left(\frac{\partial\psi^*}{\partial t}{\bf w}
         - \frac{\partial\psi}{\partial t}{\bf
         w}^*\right).\label{u1}
\end{eqnarray}
Using the Schr\"odinger equation,
\begin{eqnarray}
  i\hbar\partial_t\psi({\bf r},t) = \left[-\frac{\hbar^2}{2m}\Delta +
  V({\bf r})\right]\psi({\bf r},t),
\end{eqnarray}
we obtain
\begin{eqnarray}
  {\bf u} = \frac{\hbar}{2mi}
  \frac{{\bf w}\times{\bf w}^*}{\vert{\bf w}\times{\bf w}^*\vert^2}
    \times\left({\bf w}\Delta\psi^*+{\bf
    w}^*\Delta\psi\right).\label{u2}
\end{eqnarray}
Let us notice that the potential term $V$ does not appear in the
formula for the velocity ${\bf u}$ since the wave function
vanishes on the vortex line. The motion of the vortex line at a
given point is completely determined by the local properties of
the wave function: its gradient and its Laplacian at this point.
There is no direct relation between the velocity ${\bf u}$ of the
vortex line and the velocity ${\bf v}$ of the probability fluid.

\section{General method of generating vortex lines}\label{general}

Let us choose the initial condition for the solution of the
time-dependent Schr\"odinger equation in the form
\begin{eqnarray}\label{gen_fun}
 \phi_{\bf k}({\bf r}) = \exp(i{\bf k}\!\cdot\!{\bf r})\phi_0({\bf r}).
\end{eqnarray}
By differentiating $\phi_{\bf k}$ with respect to the components of the
vector ${\bf k}$ any number of times we obtain new wave functions. Each
differentiation brings down a component of the position vector and in
this manner we may generate an arbitrary complex polynomial that
multiplies the initial wave function. Carrying out the
differentiations, adding the results with appropriate complex
coefficients, and setting ${\bf k}=0$ at the end, we arrive at the
expression for the initial wave function of the form
\begin{eqnarray}\label{init_poly}
  (W_R({\bf r}) + i W_I({\bf r}))\phi_0({\bf r}),
\end{eqnarray}
where $W_R$ and $W_I$ are real polynomials in the three variables $x,
y$, and $z$.

The conditions $W_R({\bf r})=0$ and $W_I({\bf r})=0$ define two
two-dimensional surfaces in the three-dimensional space. When these
surfaces intersect, they define a line. This will be a vortex line if
the circulation around it is different from zero, which is the generic
case. This procedure leads to wave functions that have initially vortex
lines of almost arbitrary shape and topology embedded in them. In other
words, the wave function $\phi_{\bf k}$ serves as the generating
function for vortex lines. In practice we are facing technical
limitations; for higher order polynomials the analysis becomes very
cumbersome. However, if the polynomial $W_R + iW_I$ is constructed as a
product of some simple polynomials in the form
\begin{eqnarray}\label{product}
  W_R(x,y,z) &+& iW_I(x,y,z) = (W'_R(x,y,z) + i W'_I(x,y,z))\nonumber\\
  &\times&(W''_R(x,y,z) + i W''_I(x,y,z))\dots,
\end{eqnarray}
then the initial wave function is easily analyzed: it has several
vortex lines, each corresponding to one of the factors in
(\ref{product}). Most of our discussion will be restricted
to second degree polynomials which already lead to a
plethora of different shapes and behaviors.

The time evolution of vortex lines that were embedded in the initial
wave function is determined, of course, by the Schr\"odinger equation.
The advantage of using the generating function (\ref{gen_fun}) is that
once we find the time-dependent solution $\psi_{\bf k}({\bf r},t)$
satisfying the initial condition
\begin{eqnarray}\label{init}
  \psi_{\bf k}({\bf r},t=0) = \phi_{\bf k}({\bf r}),
\end{eqnarray}
we may use it to generate all solutions that at $t=0$ have the form
(\ref{init_poly}). Since the vector ${\bf k}$ does not appear in the
Schr\"odinger equation, the differentiations with respect to its
components do not interfere with the time evolution: ${\bf
k}$-derivatives of the time-dependent wave function are also solutions
of the Schr\"odinger equation. Carrying out the differentiations with
respect to the components of ${\bf k}$ on the time dependent function
$\psi_{\bf k}({\bf r},t)$, we shall obtain the time evolution of vortex
lines from the time dependence of the polynomials $W_R({\bf r},t)$ and
$W_I({\bf r},t)$ introduced in the formula (\ref{init_poly}).

Obviously, wave functions are not restricted to those that have only a
finite number of vortex lines described by a polynomial prefactor
(\ref{init_poly}). For example, replacing the sum of polynomials by the
sum of trigonometric functions $\cos(x/a)+i\cos(y/b)$ leads to an
infinite ``forest'' of vortex lines pointing in the $z$-direction.

\section{Vortex lines for a freely moving particle}\label{free}

In this Section we shall describe the time evolution of vortex lines
for a freely moving particle. The simplest solution of the
corresponding time-dependent Schr\"odinger equation, that has at $t=0$
the form (\ref{gen_fun}), corresponds to $\phi_0({\bf r})=1$,
\begin{eqnarray}\label{plane_wave}
\psi_{\bf k}({\bf r},t) = \exp(i{\bf k}\!\cdot\!{\bf r})
\exp(-i\hbar{\bf k}^2t/2m).
\end{eqnarray}
Such a plane-wave solution is not square-integrable but that can easily
be corrected by multiplying (\ref{plane_wave}) by a slowly varying
envelope without affecting significantly the dynamics of vortex lines.
We shall return to this problem at the end of this Section.

Let us begin with the simplest solution that initially has one
rectilinear vortex described by the wave function $(x\cos\chi + i
y\sin\chi)\exp(i{\bf k}\!\cdot\!{\bf r})$. The parameter $\chi$
measures the degree of squeezing of the velocity field.
Differentiating the wave function (\ref{plane_wave}) with respect
to $k_x$ and $k_y$ we obtain the wave function with a vortex line.
Since the parameters $k_x,k_y$, and $k_z$ in (\ref{plane_wave})
can be interpreted as the components of the particle momentum,
$\hbar{\bf k}={\bf p}$, it is instructive not to set ${\bf k}$ to
zero but keep it finite after the differentiation. This will
result in a free motion of the whole vortex line with the velocity
${\bf v} = \hbar{\bf k}/m$:
\begin{eqnarray}\label{line_vor}
  \left(x(t)\cos\chi + i y(t)\sin\chi\right)\psi_{\bf k}({\bf r},t),
\end{eqnarray}
where $x(t)=x-v_x t$ and $y(t)=y-v_y t$. Throughout this paper
$x(t), y(t)$ and $z(t)$ will always denote a free classical motion
with constant velocity. Eq. (\ref{line_vor}) means that the vortex
line riding on the plane wave is moving uniformly, without
changing its shape, with the speed determined by the particle
momentum. This was to be expected as a consequence of the
Gallilean invariance of the theory. The same free motion of a
single vortex line was found in the nonlinear Schr\"odinger
equation \cite{fetter1}.

\begin{figure}
  \begin{center}
    \epsfbox{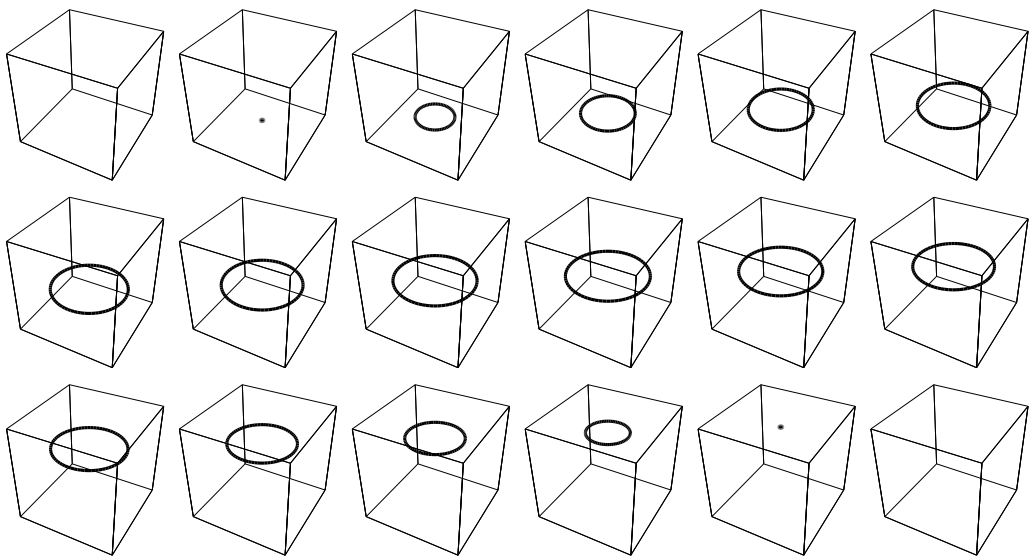}
  \end{center}
\caption{Creation of a vortex at a point. A vortex ring is created at a
point where the wave function vanishes.}\label{fig:vor_creat1}
\end{figure}

Our next example of a vortex is a ring of radius $R$. It can be
obtained, for example, by cutting a cylinder by a plane:
\begin{eqnarray}\label{line_vor2}
  W_R(x,y,z) = x^2 + y^2 - R^2,\,\,\,\,
  W_I(x,y,z) = a z.
\end{eqnarray}
The ratio of the two dimensional parameters $R$ and $a$ determines the
degree of squeezing of the vortex, $\tan\chi = a/2R$. The motion of
this vortex is described by the time-dependent wave function
\begin{eqnarray}\label{circle_vor1}
\psi^C({\bf r},t)\!
 &=&\!\left[x(t)^2+y(t)^2-R^2+ia\!\left(\!z(t) +
  \frac{2\hbar t}{ma}\!\right)\right]\nonumber\\
  &\times&\psi_{\bf k}({\bf r},t).
\end{eqnarray}
The quantum correction (the term $2\hbar t/ma$) to the classical
vortex motion that appears here amounts only to a change of the
vortex velocity in the direction of the cylinder axis. In this
example, the vortex ring in the form of a circle has been produced
by intersecting a cylinder and a plane. However, there are many
other possibilities, even with the use of only simple surfaces. It
is easy to obtain at $t=0$ the same shape of the vortex line
choosing different wave functions. The motion of these vortices
may change dramatically, when one pair of surfaces is replaced by
a different one. The circular vortex formed by intersecting a
cylinder with a plane, as shown above, moves uniformly along the
cylinder axis. On the other hand, the same vortex produced at
$t=0$ by a sphere and a plane will contract to a point and
disappear. The appropriate wave function has the form
\begin{eqnarray}\label{circle_vor2}
\psi^S({\bf r},t)
 &=&\left[x(t)^2+y(t)^2+z(t)^2-R^2+ia\!\left(\!z(t)+
  \frac{3\hbar t}{ma}\!\right)\right]\nonumber\\
  &\times&\psi_{\bf k}({\bf r},t).
\end{eqnarray}
The circle disappears, i.e., the vortex is annihilated, at the
time $t_a = maR/3\hbar$. Extending our analysis to negative values
of $t$, we see that this vortex was created at the time $t_c =
-maR/3\hbar$ (Fig.~\ref{fig:vor_creat1}). A circular vortex
obtained by intersecting a paraboloid with a plane has again a
different motion: after it is created, it keeps expanding and
moving along the $z$-axis to infinity. Other motions of a circular
vortex line can also be obtained by choosing a second quadratic
surface, instead of a plane. In all these cases the motion of the
vortex ring has two components. One is the free motion with the
velocity ${\bf v}$ and the other is a quantum correction that may
lead to the vortex annihilation. We would like to stress that the
creation and annihilation of vortex rings are of a purely quantum
origin. These phenomena are not found in the studies of motion of
the vortex ring based on the semiclassical approximation
\cite{fetter1,fetter2}.

The creation and annihilation of vortex rings at a point, shown in
Fig.~\ref{fig:vor_creat1}, does not contradict the Helmholtz-Kelvin
theorem on the conservation of circulation. This theorem, obviously,
holds only inside the fluid but the vortex creation and the vortex
annihilation occur at the points where the fluid density vanishes.

\begin{figure}
  \begin{center}
    \epsfbox{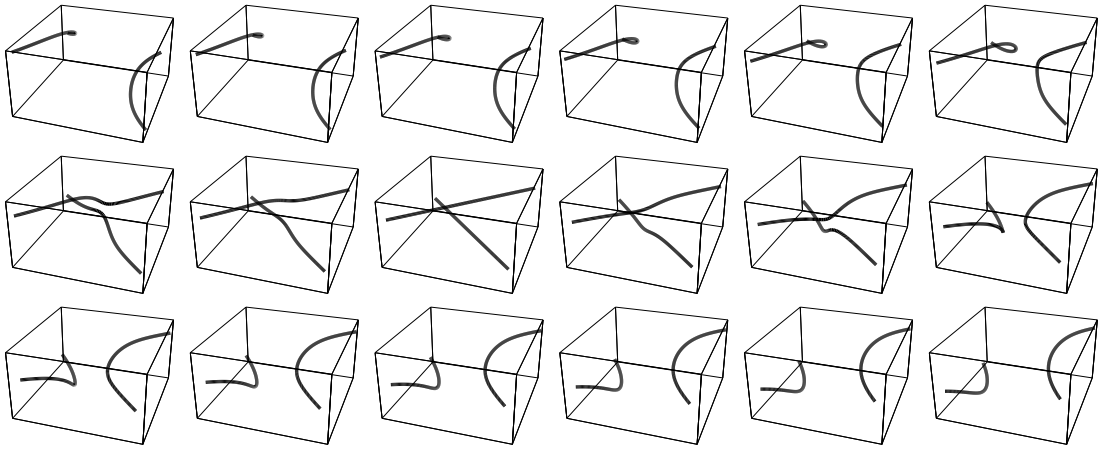}
  \end{center}
\caption{Two vortex lines evolving in time from negative to positive
values of $t$. At $t=0$ these vortex lines form a pair of
nonintersecting straight lines.}\label{fig:two_vor}
\end{figure}

The mutual interaction between vortex lines can be seen in the
motion of two vortex lines. The initial wave function containing
two arbitrary rectilinear vortices has the product form
\begin{eqnarray}\label{two_vor}
 \phi_{\bf k}({\bf r}) = ({\bf w}_1\!\cdot\!({\bf r}-{\bf r}_1))
 ({\bf w}_2\!\cdot\!({\bf r}-{\bf r}_2))\exp(i{\bf k}\!\cdot\!{\bf r}).
\end{eqnarray}
The corresponding time-dependent wave function, obtained by our method
is
\begin{eqnarray}\label{two_vor1}
 \psi({\bf r},t)&=&[({\bf w}_1\!\cdot\!({\bf r}(t)-{\bf r}_1))
 ({\bf w}_2\!\cdot\!({\bf r}(t)-{\bf r}_2))+
 \frac{i\hbar t}{m}{\bf w}_1\!\cdot\!{\bf w}_2]\nonumber\\
 &\times&\psi_{\bf k}({\bf r},t).
\end{eqnarray}
Thus, the two vortices become, in general, entangled by the
quantum correction term: the wave function is not a product of
factors describing two separate vortices. The coupling between the
two vortices is due to the interaction between the vortices and it
vanishes only when ${\bf w}_1\!\cdot\!{\bf w}_2=0$. The time
evolution exhibiting the phenomenon of a switchover of the two
vortex lines is shown in Fig.~\ref{fig:two_vor}. To simplify the
calculations we assumed that both vortices are nonsqueezed. The
coordinate system is chosen in such a way that at $t=0$ the vortex
lines lie in the $z=\pm a$ planes with the angle between them
equal to $2\varphi$. With this choice, the time-dependent wave
function can be written in the form
\begin{eqnarray}\label{two_vor2}
 \psi({\bf r},t) =\left[W_1 W_2 - (2i\hbar t/m)\sin^2\varphi\right]
 \psi_{\bf k}({\bf r},t),
\end{eqnarray}
where the two polynomials $W_1$ and $W_2$,
\begin{eqnarray}\label{two_vor3}
 W_1({\bf r},t) = x(t)\cos\varphi + y(t)\sin\varphi
  + i(z(t)+a),\\
 W_2({\bf r},t) = x(t)\cos\varphi - y(t)\sin\varphi
  + i(z(t)-a),
\end{eqnarray}
describe the motion of both vortex lines separately, whereas the
last term in the square bracket is responsible for their
interaction. Only when $\varphi=0$ the vortex lines remain
unchanged during the time evolution. This behavior is different
from what can be found in non-compressible fluids \cite{lamb}. The
quantum interaction between the vortex lines is the strongest when
the lines are antiparallel (i.e. the two vortices have opposite
circulation, $\varphi=\pi/2$). In this case the vortex lines
retain their direction but the distance between them decreases and
at time $t_a=ma^2/\hbar$ the vortex lines collide and annihilate.
The same wave function describes at $t_c=-ma^2/\hbar$ the creation
of two rectilinear vortices of opposite circulation. These
processes are shown in Fig.~\ref{fig:vor_creat2}.

We shall now replace the plane wave solution of the Schr\"odinger
equation, corresponding to $\phi_0({\bf r})=1$, by a normalizable
Gaussian wave packet $\phi_0({\bf r})=\exp(-{\bf r}^2/2l^2)$. This will
give us square integrable solutions with vortex lines. The generating
function for such solutions has the form:
\begin{eqnarray}\label{gaussian}
 \psi_{\bf k}^G({\bf r},t) = \frac{\exp(-{\bf k}^2l^2/2)}
 {(1 + i\hbar t/ml^2)^\frac{3}{2}}
 \exp\!\left[\frac{-({\bf r} -i{\bf k}l^2)^2}
 {2l^2(1 + i\hbar t/ml^2)}\right].
\end{eqnarray}
We shall illustrate the influence of the shape of the Gaussian
envelope on the motion of vortices in the simplest case of one
rectilinear nonsqueezed vortex that is parallel to the $z$-axis
and passes through the point $(x_0,0)$ at $t=0$. The motion of
such a line is composed of the free motion with the velocity ${\bf
v}$ of the center of the Gaussian envelope and the motion induced
by the shape of the envelope. The time-dependent wave function in
this case is
\begin{eqnarray}\label{gauss_vor}
 \frac{x(t)-x_0 + i(y(t)-x_0\hbar t/ml^2)}
 {1 + i\hbar t/ml^2}\psi_{\bf k}^G({\bf r},t).
\end{eqnarray}
The last term in the numerator represents the quantum correction to the
free vortex motion and it vanishes in the limit, when the width of the
Gaussian envelope $l$ tends to infinity. The speed of this additional
motion depends on the smallest distance of the vortex line from the
center of the Gaussian but does not depend on the motion of its center.

\section{Vortex lines for a particle in a uniform magnetic field}
\label{magnetic}

\begin{figure}
  \begin{center}
    \epsfbox{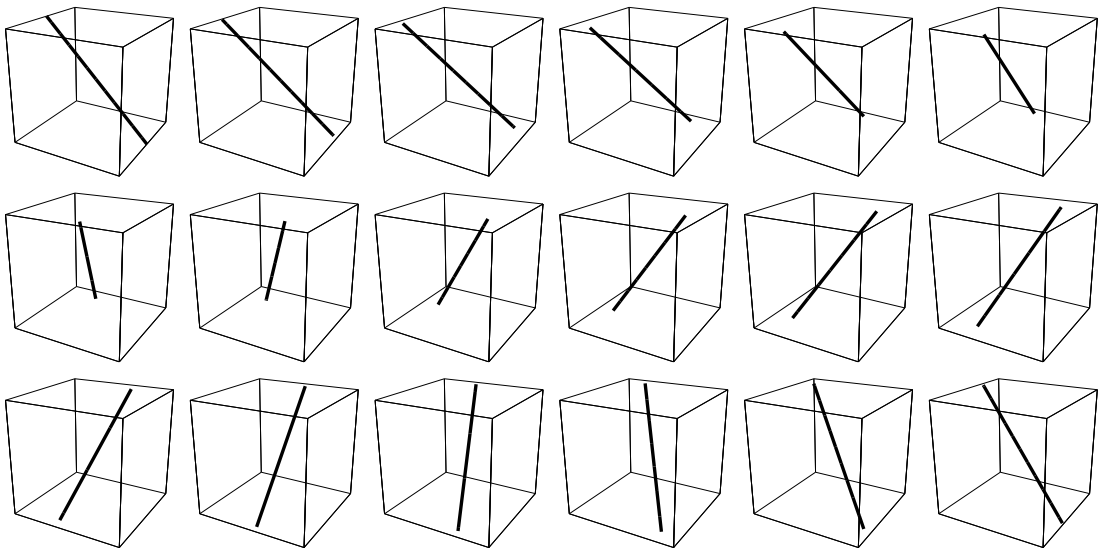}
  \end{center}
\caption{Precession of a rectilinear vortex in a magnetic
field.}\label{fig:mag_line}
\end{figure}

In order to determine the motion of vortex lines in a magnetic field we
need the counterpart of the generating function (\ref{plane_wave}) in
the presence of the field. We shall use the Schr\"odinger equation in
the symmetric gauge
\begin{eqnarray}\label{sch_mag}
 &&i\partial_t\psi({\bf r},t)=\big[-\frac{\hbar^2}{2m}\Delta
 \nonumber\\
 && - \frac{i\hbar e B}{m}(x\partial_y-y\partial_x)
  + \frac{e^2B^2}{8m} (x^2+y^2)\big]\psi({\bf r},t),
\end{eqnarray}
where the $z$-axis is taken along the magnetic field. The solution
of this equation, that at the time $t=0$ has the form $\exp(i{\bf
k}\!\cdot\!{\bf r}) \exp(-eB(x^2+y^2)/4\hbar)$ (a plane wave
multiplied by the ground state Gaussian in the $xy$-plane), is
\begin{eqnarray}\label{gen_mag}
 \psi_{\bf k}^M({\bf r},t)&=&\exp(-eB(x^2+y^2)/4\hbar)\exp[-i\omega_ct/2]\nonumber\\
 &\times&\exp[\hbar(e^{-i\omega_ct}-1)(k_x^2+k_y^2)/(2eB)]\nonumber\\
 &\times&\exp[i(e^{-i\omega_ct}+1)(xk_x+yk_y)/2]\nonumber\\
 &\times&\exp[(e^{-i\omega_ct}-1)(xk_y-yk_x)/2]\nonumber\\
 &\times&\exp[izk_z-i\hbar k_z^2/(2eB)],
\end{eqnarray}
where $\omega_c=eB/m$ is the cyclotron frequency. Since the
differentiation with respect to the components of ${\bf k}$ brings
down linear terms in the space variables, a single vortex along a
straight line will preserve its form. All that can happen is the
motion of this straight line. For example, a line vortex that at
$t = 0$ lies in the $y = a$ plane at an angle $\varphi$ to the
field direction will move in time according to the following
parametric representation
\begin{eqnarray}\label{line_mag}
y(x) &=& \frac{2a + x\sin(\omega_ct)
      \left(1 + \sin{\varphi}\right)}
      {1 - \sin{\varphi} + \left(1 + \sin{\varphi}\right)\cos(\omega_ct)},\\
z(x) &=& -\frac{2x\tan{\varphi} +
      a\sin(\omega_ct)\left(\sec{\varphi} + \tan{\varphi}\right)}
         {1 - \sin{\varphi} + \left(1 + \sin{\varphi}\right)\cos(\omega_ct)}.
\end{eqnarray}
In Fig.~\ref{fig:mag_line} we show the motion of this line during one
period of the cyclotron motion.

\section{Vortex lines for a particle in a trap}\label{trap}

\begin{figure}
  \begin{center}
    \epsfbox{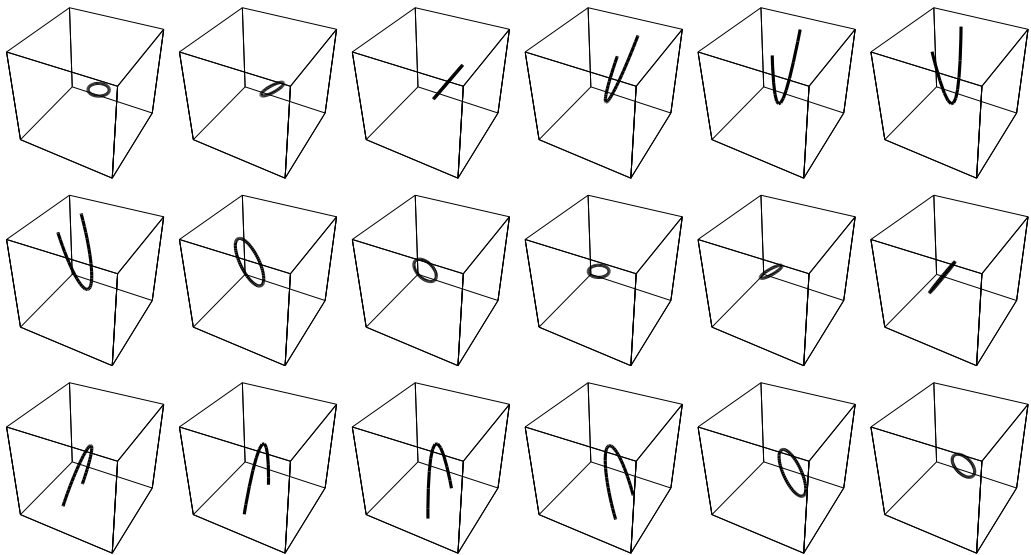}
  \end{center}
\caption{Evolution of a vortex ring in a harmonic trap.}\label{fig:vor_trap}
\end{figure}

The study of vortex dynamics in a trap has a special significance
because it may, perhaps, throw some light (as the first, linear
approximation) on the behavior of the atoms that form the
Bose-Einstein condensate. For simplicity, we shall consider a trap
in the form of a spherically symmetric harmonic oscillator with
the frequency $\omega$ and choose as $\phi_0({\bf r})$ in the
formula (\ref{gen_fun}) the ground state wave function
$\exp(-m\omega{\bf r}^2/2\hbar)$. The time dependent wave function
that serves as a generator of vortex lines in this case is
\begin{eqnarray}\label{gen_trap}
\psi_{\bf k}^T({\bf r},t)
 &=&\exp(-3i\omega t/2)\exp(-\frac{m\omega}{2\hbar}{\bf r}^2)\nonumber\\
&\times&\exp\left(ie^{-i\omega t}({\bf k}\!\cdot\!{\bf r}
-\frac{\hbar{\bf k}^2}{2m\omega}\sin(\omega t))\right).
\end{eqnarray}
As an example we shall choose the vortex line that at $t=0$ has
the form of a circular ring of radius $R$ passing through the
center. The wave function describing the time evolution of such a
vortex is obtained by evaluating the appropriate derivatives of
(\ref{gen_trap}). Setting ${\bf k}=0$ after the differentiation,
we obtain
\begin{eqnarray}\label{ring_trap}
\big[e^{-2i\omega t}(x^2+y^2-\frac{\hbar}{m\omega})
+\frac{\hbar}{m\omega}&-& e^{-i\omega t}R(2x-iz)\big]\nonumber\\
 &\times&\psi_{\bf k=0}^T({\bf r},t).
\end{eqnarray}
The time evolution of this vortex line is shown in
Fig.~\ref{fig:vor_trap}.

\section{Vortex lines in relativistic wave mechanics}\label{rel}

The description of vortex lines and their motion presented in this
paper is not restricted to nonrelativistic wave mechanics based on the
Schr\"odinger equation. It can be extended to a relativistic theory
based on the Klein-Gordon equation. Let us consider a relativistic wave
function $\phi$ of a free particle in the form of a plane wave
\begin{eqnarray}\label{rel_pw}
 \phi_{\bf k}({\bf r},t)=\exp(i{\bf k}\!\cdot\!{\bf r})\exp(-i\omega_{\bf k}t),
\end{eqnarray}
where $\omega_{\bf k}=c\sqrt{{\bf k}^2+(mc/\hbar)^2}$. This
solution, in full analogy with the nonrelativistic case, may serve
as a generating function for solutions with all kinds of vortex
lines. For example, by taking first derivatives we generate a
solution with one rectilinear vortex
\begin{eqnarray}\label{rel_vor1}
 \left(x(t)\cos\chi + i y(t)\sin\chi\right)\phi_{\bf k}({\bf r},t).
\end{eqnarray}
The only difference between this solution and the nonrelativistic one
(\ref{line_vor}) is due to a different relation between the velocity
and the wave vector. In the relativistic case it is ${\bf v} =
\nabla_{\bf k}\omega_{\bf k} = \hbar{\bf k}/\sqrt{(\hbar{\bf k}/c)^2 +
m^2}$, instead of just $\hbar{\bf k}/m$. Otherwise, the result is the
same: the vortex line moves with the constant velocity ${\bf v}$. The
motion of two vortex lines can also be determined in the same way as in
the nonrelativistic case. In the limit, when ${\bf k} \to 0$ the
relativistic case reduces to the nonrelativistic case,
Eq.(\ref{two_vor2}).

With the help of the generating function we may also determine the
motion of a vortex ring in the relativistic theory. The wave function
that determines the motion of the vortex ring in the relativistic case
(the counterpart of the nonrelativistic wave function
(\ref{circle_vor1})) is
\begin{eqnarray}\label{circle_vor3}
\psi^C({\bf r},t)
 &=& \Bigl[x(t)^2 + y(t)^2 - R^2\nonumber\\ + i a\bigl(z(t) &-&
  \frac{2\hbar(1 - c^2(k_x^2+k_y^2)/\omega^2)}
  {a\sqrt{(\hbar{\bf k}/c)^2+m^2}}t\bigr)\Bigr]\phi_{\bf k}({\bf r},t).
\end{eqnarray}
The speed of the vortex depends on the arbitrary parameter $a$ and may
exceed the speed of light when $a$ is sufficiently small. This
phenomenon, however, does not contradict the theory of relativity. It
is quite similar to the so called superluminal propagation in optics
\cite{cw,cks}, when the maximum of the signal travels with the velocity
greater than $c$. In the present case it is the node of the probability
density that may travel at an arbitrarily high speed.

The presence of vortex lines is not restricted to relativistic
wave functions with only one component. They can also be found in
wave functions with several complex components describing spinning
particles. However, the freedom of attaching almost arbitrary
vortex lines to any wave function is gone. There are stringent
condition that must be satisfied in order to have physically
admissible solutions. For example, a rectilinear vortex can be
attached to a plane wave solution of the Dirac equation, but only
when the direction of the vortex line is correlated with the
particle momentum.

Of particular interest are the vortex lines associated with photon
wave functions. Simple states of that category are the well known
multipolar states: the eigenstates of the angular momentum. The
corresponding photon wave functions have rectilinear vortices
running from $-\infty$ to $\infty$, as in nonrelativistic wave
mechanics. More elaborate vortex lines for photon wave functions
are also possible but their study falls outside the scope of the
present paper.

\section{Conclusions}\label{concl}

In our study of the motion of vortex lines in wave mechanics we
have shown that this motion is determined by four elements: the
shape of the vortex line, the shape of the wave function, the
interaction between different vortex lines, and the external
forces acting on the particle. Complete isolation of these effects
is not possible but one can gain some insight by appropriate
limiting procedures. We found that quantum corrections to the
vortex motion are responsible for the mutual interaction between
vortex lines and also for the vortex creation and annihilation. We
believe that the qualitative features of the vortex motion will
remain valid even in the much more complicated case of the
mutually interacting particles, when the linear Schr\"odinger
equation must be replaced by the nonlinear one.

\section*{Acknowledgement}

This work has been supported by the KBN Grant 2P30B04313.

\end{multicols}
\newpage

\end{document}